\documentclass[fleqn,12pt]{wlscirep}
\usepackage[T1]{fontenc}
\usepackage{bm}
\usepackage{float}
\DeclareUnicodeCharacter{2212}{-}

\title{Dynamics of Collective Modes in an unconventional Charge Density Wave system BaNi$_{2}$As$_{2}$}

\author[1,*]{Amrit Raj Pokharel}
\author[1,*]{Vladimir Grigorev}
\author[2]{Arjan Mejas}
\author[1]{Tao Dong}
\author[3]{Amir A. Haghighirad}
\author[3]{Rolf Heid}
\author[3]{Yi Yao}
\author[3]{Michael Merz}
\author[3]{Matthieu Le Tacon}
\author[1,**]{Jure Demsar}

\affil[1]{Institute of Physics, Johannes Gutenberg University, 55128 Mainz, Germany}
\affil[2]{Institute of Solid State Physics, TU Wien, 1040 Vienna, Austria}
\affil[3]{Institute for Quantum Materials and Technologies, Karlsruhe Institute of Technology, 76344 Karlsruhe, Germany}
\affil[**]{e-mail: demsar@uni-mainz.de}

\begin{abstract}
BaNi$_{2}$As$_{2}$ is a non-magnetic analogue of BaFe$_{2}$As$_{2}$, the parent compound of a prototype pnictide high-temperature superconductor, displaying superconductivity already at ambient pressure. Recent diffraction studies demonstrated the existence of two types of periodic lattice distortions above and below the triclinic phase transition, suggesting the existence of an unconventional charge-density-wave (CDW) order. The suppression of CDW order upon doping results in a sixfold increase in the superconducting transition temperature and enhanced nematic fluctuations, suggesting CDW is competing with superconductivity. Here, we apply time-resolved optical spectroscopy to investigate collective dynamics in BaNi$_{2}$As$_{2}$. We demonstrate the existence of several CDW amplitude modes. Their smooth evolution through the structural phase transition implies the commensurate CDW order in the triclinic phase evolves from the high-temperature unidirectional incommensurate CDW, and may indeed trigger the structural phase transition. Excitation density dependence reveals exceptional resilience of CDW against perturbation, implying an unconventional origin of the underlying electronic instability.  

\end{abstract}
\begin{document}

\flushbottom
\maketitle
\def\thefootnote{*} 
\footnotetext{These authors contributed equally to this work}
\def\thispagestyle{empty}
\section*{Introduction}
As in cuprate superconductors, high-temperature superconductivity in Fe-based
superconductors \cite{First,FeSe,LiFeAs,NaFeAs,NaFeCoAs} is found in the
proximity to the magnetically ordered state, with the interplay between the
magnetic order and superconductivity long being at the forefront of the
research. Detailed studies soon revealed the presence of another type of
ordering, the electronic nematicity, where the degeneracy between the two
equivalent orthogonal directions in the square Fe planes is lifted, inducing a
symmetry reduction at $T_{\mathrm{S}}$ from tetragonal to orthorhombic.\cite{Chu,Joerg}
This so-called nematic phase transition, which in most cases precedes the
magnetic one, has been considered to be driven by magnetic fluctuations.
Indeed, in the parent compound of the prototypical system, BaFe$_{2}$As$_{2}$
(Fe-122), a stripe-type spin-density-wave ground state is realized below
$T_{\mathrm{M}}\approx$ 134 K, the transition being slightly preceded by the nematic
one at $T_{\mathrm{S}}\approx$ 137 K.\cite{MGKim} Upon doping (or application of
pressure) high-temperature superconductivity is realized in this system.

BaNi$_{2}$As$_{2}$ (Ni-122) is a non-magnetic analogue of Fe-122. It shares
the same tetragonal high-temperature structure while below $T_{\mathrm{S}}$ = 138 K
the structure is triclinic. It displays superconductivity
already in the undoped case ($T_{\mathrm{c}}\approx0.6$ K) with no magnetic order
reported down to the lowest temperatures.\cite{Ronning} Instead, recent X-ray
diffraction studies suggest two distinct charge-density-wave (CDW) orders
above and below $T_{\mathrm{S}}$.\cite{XRD} What is particularly striking, is the
observed six-fold enhancement of the superconducting $T_{\mathrm{c}}$ and a giant
phonon softening observed when doping Ni-122 to a level where\ the structural
transition is completely suppressed.\cite{Kudo} Similar $T_{\mathrm{c}}$ enhancement at
the possible quantum critical point between the triclinic and tetragonal
phases was recently observed also in strontium substituted
Ni-122.\cite{Paglione} There, electronic nematic fluctuations were
demonstrated and showed a dramatic increase near the suggested quantum
critical point.\cite{Paglione} The observed correlation between the
enhancement of superconductivity and the increase in nematic fluctuations,
with the same B$_{1g}$ symmetry breaking for both the nematic fluctuations and
the CDW order, may suggest a charge-order-driven electronic nematicity in
Ni-122.\cite{Paglione} The interplay between electronic nematicity, CDW order
and superconductivity in Ni-122 system thus presents one of the key topics in
the current pnictide research, especially given the parallels to cuprate
superconductors\cite{Cuprates} that can be drawn.

The existence of the periodic lattice distortion (PLD) in Ni-122 system has
been demonstrated by X-ray diffraction studies.\cite{XRD,Paglione,Lee21,Merz} In undoped Ni-122 diffuse incommensurate superstructure reflections at ($h\pm0.28,k,l$) are observed already at room temperature\cite{Merz} (the
indexing throughout of the paper refers to the high-\textit{T} tetragonal
phase). Upon cooling, the correlation length of modulation strongly increases around 150K, yet the system remains tetragonal.\cite{Lee21,Merz} Thermal expansion studies\cite{Merz} reveal a second order phase transition at $T_{\mathrm{S}'} \approx 142$ K, where orthorhombic distortion implies an unidirectional incommensurate CDW, I-CDW$_{1}$, at ($h\pm0.28,k,l$). At $T_{\mathrm{S}}\approx 138$ K\cite{XRD} a first order structural phase transition to a triclinic structure takes place (both $T_{\mathrm{S}}$ and $T_{\mathrm{S}'}$ are transition temperatures upon warming). In the triclinic phase, a new periodicity of PLD is observed, attributed to I-CDW$_{2}$ with ($h\pm
1/3+\delta$, $k,$ $l\mp1/3+\delta$) superstructure reflections. The
discommensuration vanishes, \textit{i.e.} $\delta\rightarrow0$, slightly below
$T_{\mathrm{S}}$, resulting in a commensurate C-CDW\cite{XRD,Lee21,Merz} with a wave-vector
($1/3$, $0,1/3$). There are, however, no abrupt changes in the displacement
amplitude at the lock-in transition.\cite{XRD} Given the fact that the
triclinic transition is of the first order, with hysteresis of about 5 K, one
can argue that the structural transition is concomitant with the I-CDW$_{1}$
to C-CDW transition.

In the charge channel, however, optical studies show no signatures of the
CDW-induced optical gap.\cite{WangPRB,Wang} To get further support for the CDW
origin of the observed PLD\cite{XRD,Paglione,Merz} and to gain insights into
the relation between CDWs and structure, the information on CDW collective
modes is required. To this end, we apply time-resolved optical spectroscopy
which has been demonstrated to be particularly sensitive to study low-energy
\textbf{q}$\approx0$ Raman-active collective modes in systems exhibiting CDW
order.\cite{TaSTaSe,Scha10,TSI,Scha14} While obeying similar selection rules
to conventional Raman spectroscopy,\cite{Stevens} the method was able to
spectrally resolve modes with line widths as low as 3 GHz\ ($0.1$ cm$^{-1}%
$),\cite{Scha10} access modes at frequencies down to 100 GHz (3 cm$^{-1}%
$),\cite{TSI} and is suitable also for investigations of
disordered/inhomogeneous samples.\cite{Dominko1} The information on the
temperature ($T$)\cite{TaSTaSe,Scha10,TSI,Scha14} and excitation fluence ($F$)
dependent\cite{Tome09,Yusupov,Stojchevska} dynamics provide further insights
into the nature of collective ground states.

Here, we report systematic $T$- and $F$-dependent study of photoinduced
reflectivity dynamics in BaNi$_{2}$As$_{2}$. At temperatures below
$\approx150$ K we observe several oscillatory modes. Comparison to phonon
dispersion calculations reveals that several weak modes could be attributed to
\textbf{q}$=0$ Raman active modes. The dominant, strongly temperature dependent modes,
however, do not match the calculated \textbf{q}$=0$ phonon frequencies of the high
temperature tetragonal phase. Based on their $T$- and $F$-dependence, which
match the behavior seen in prototype CDW systems,\cite{Scha10} we attribute
these to collective amplitude modes of the CDW.\cite{Scha10,rice,Levanyuk} As
these modes appear at temperatures $\approx10$ K above the structural phase
transition and show a continuous behavior across $T_{\mathrm{S}}$, we conclude that the
in-plane I-CDW$_{1}$ transforms into the C-CDW by gaining additional
periodicity along the c-axis. This result suggests that while in Fe-122
nematicity is driven by magnetic instability, in BaNi$_{2}$As$_{2}$ the
structural transition may be driven by the CDW instability. Moreover, while
the temperature dependence of collective modes in BaNi$_{2}$As$_{2}$ follows
the behavior seen in conventional Peierls CDW systems, the behavior is
substantially different as far as the excitation density dependence is
concerned. In particular, no ultrafast collapse of the CDW is observed up to
excitation densities over an order of magnitude higher than in prototype
Peierls systems,\cite{Scha10,Scha14,Yusupov,Schmitt,Stojchevska} suggesting an
unconventional microscopic mechanism.

\section*{Results}

We studied the $T$- and $F$-dependence of the photoinduced near-infrared
reflectivity dynamics in undoped BaNi$_{2}$As$_{2}$ using an optical
pump-probe technique. The single crystals were cleaved along the $a-b$ plane
with the pump and the probe beams at near normal incidence (they were
cross-polarized for higher signal to noise ratio). We performed also pump- and
probe-polarization dependence of the photoinduced reflectivity, with no
significant variation being observed (see Supplementary Information). The
reported temperature dependence measurements were performed upon warming.
Continuous laser heating was experimentally determined to be about 3 K for $F=0.4$ mJ cm$^{-2}$, and has been taken into account (see Supplementary Information).

\subsection*{Photoinduced reflectivity dynamics in\ the near-infrared}

\begin{figure}[pth]
\centerline{\includegraphics[width=120mm]{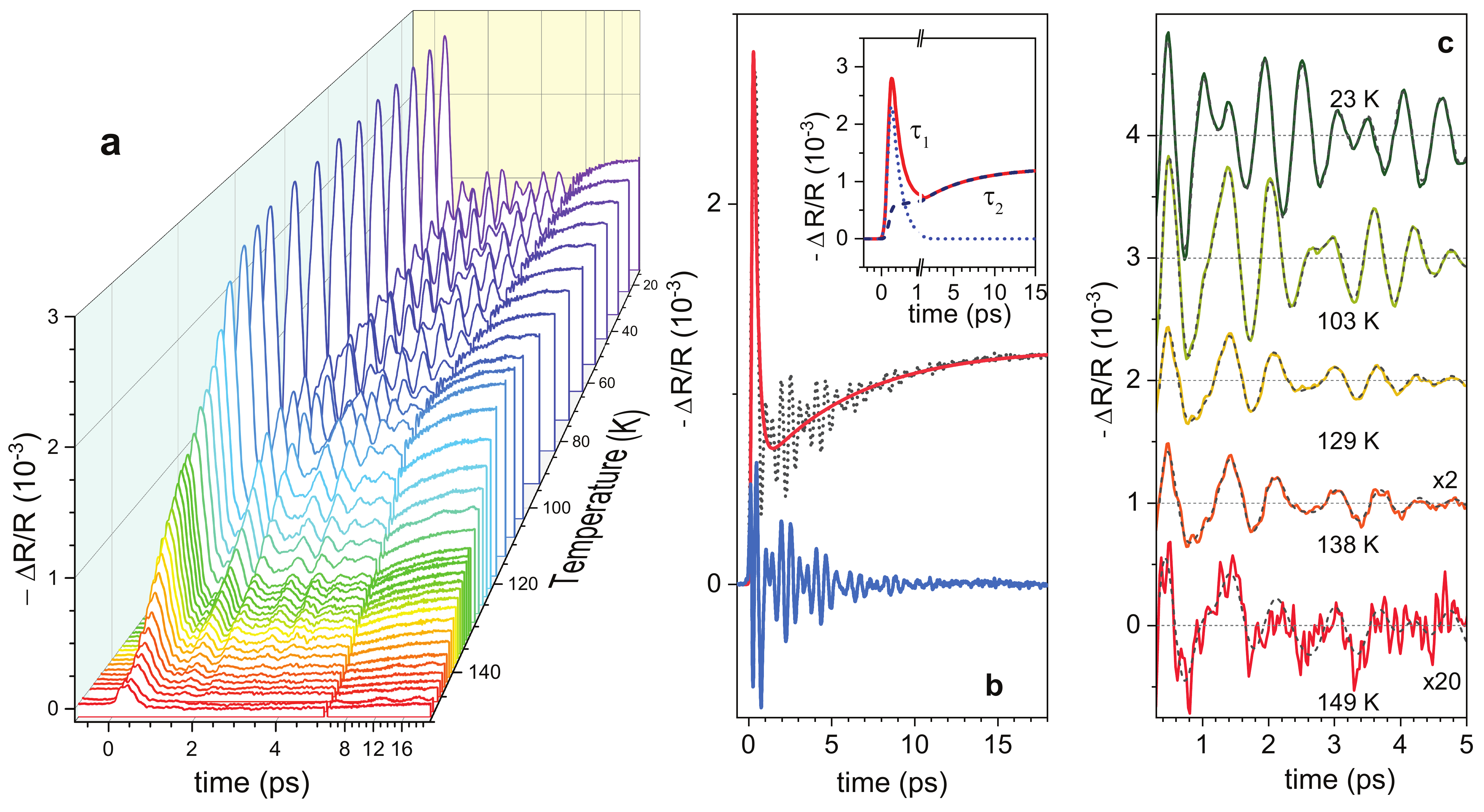}}\caption{
\textbf{Photo-induced in-plane reflectivity traces on undoped BaNi$_{2}$As$_{2}$ single crystal.}
\textbf{a} Transient reflectivity traces between 13 and 149 K, measured with $F=0.4$ mJ cm$^{-2}$ upon
increasing the temperature. \textbf{b} Decomposition of the reflectivity
transient at 13 K (black dotted line) into an overdamped (solid red line) and
oscillatory (solid blue line) components. Insert shows the individual
overdamped components (dotted blue and dash-dotted olive line). \textbf{c}
Oscillatory response at selected temperatures, together with fits using a sum of four damped oscillators (black dashed lines). Signals at 138 and 149 K are
multiplied by a factor of 2 and 20, respectively.}%
\label{Fig1}%
\end{figure}

Figure \ref{Fig1}\textbf{a} presents the \textit{T}-dependence of photoinduced
reflectivity transients, $\Delta R/R(t)$, recorded upon increasing the
temperature from 10 K, with $F=0.4$ mJ cm$^{-2}$. This fluence was chosen such
that the response is still linear, yet it enables high enough dynamic range to
study collective dynamics (see also section on excitation density dependence).

Clear oscillatory response is observed up to $\approx150$ K, with the
magnitude displaying a strong decrease near and above $T_{\mathrm{S}}$. Similarly to
the oscillatory signal, the overdamped response is also strongly
$T$-dependent. As shown in Figure \ref{Fig1}\textbf{b} the response can be
decomposed into an overdamped and oscillatory response. To analyze the
dependence of the oscillatory response on $T$, we first subtract the
overdamped components. These can be fit by
\begin{equation}
\frac{\Delta R}{R}=H\left(  \sigma,t\right)  \left[  A_{1}e^{-t/\tau_{1}%
}+B+A_{2}\left(  1-e^{-t/\tau_{2}}\right)  \right]  ,\label{Eq1}%
\end{equation}
where $H(\sigma,t)$ presents the Heaviside step function with an effective
rise time $\sigma$. The terms in brackets represent the fast decaying process
with $A_{1},\tau_{1}$ and the resulting quasi-equilibrium value $B$, together with the slower buildup
process with $A_{2}$ and $\tau_{2}$, taking place on a 10 ps timescale - see
inset to Figure \ref{Fig1}\textbf{b}. Figure \ref{Fig1}\textbf{c}, presents the oscillatory part of the signal subtracted from the overdamped response at selected temperatures together with the fit (black dashed lines) using sum of four damped oscillators (discussed below). 

\subsection*{Collective modes in BaNi$_{2}$As$_{2}$}

\begin{figure}[pth]
\centerline{\includegraphics[width=120mm]{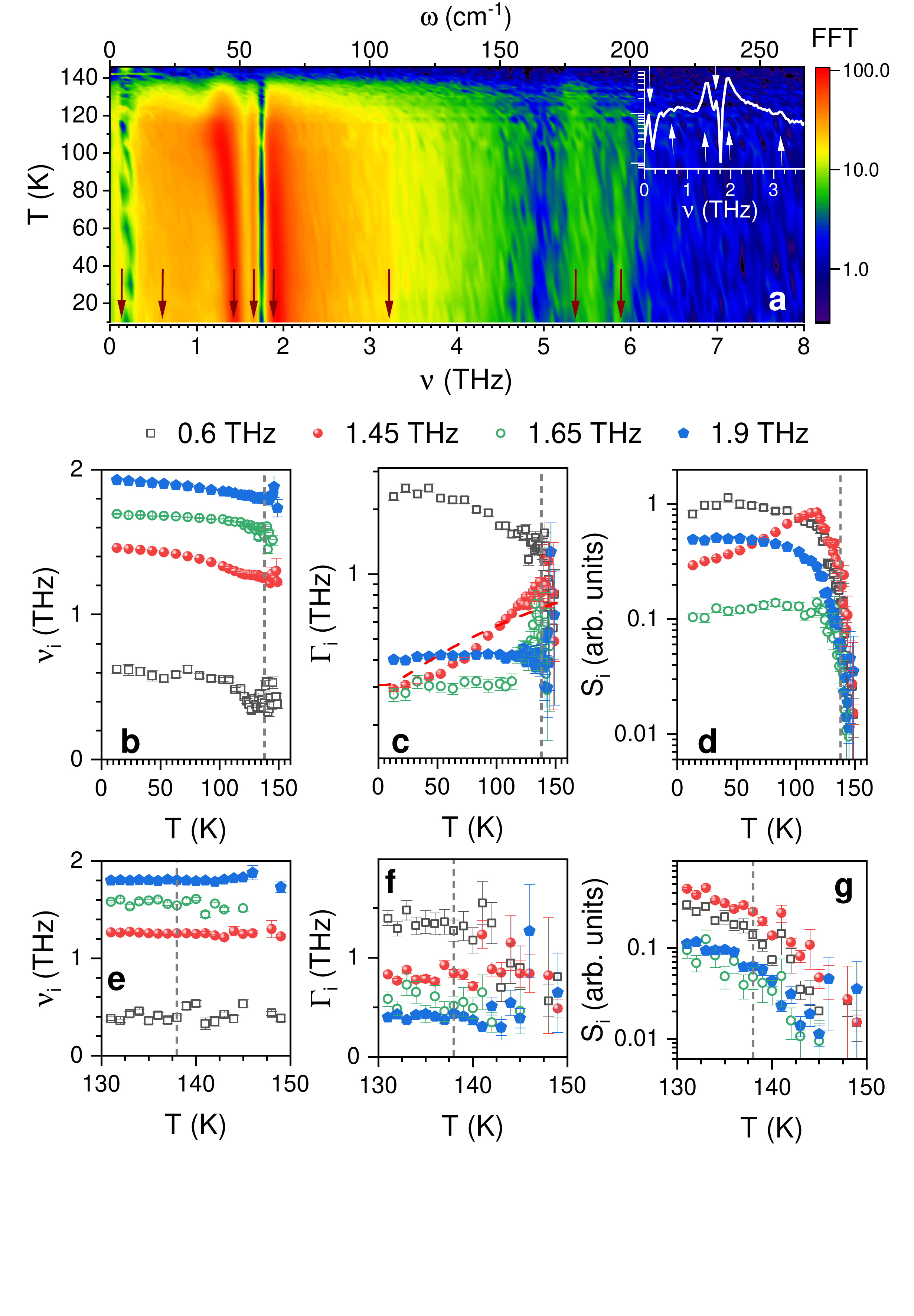}}\caption{\textbf{Analysis of
the oscillatory response.} \textbf{a} Temperature dependence of the FFT spectra, demonstrating the presence of several modes at low
temperatures. The extracted mode frequencies in the low temperature limit are denoted by red arrows (see also Supplementary Information). Insert presents the FFT of the data recorded at 13 K, with white arrows pointing at the modes. \textbf{b}-\textbf{d} The temperature dependence of the parameters of the four
strongest low-frequency modes, obtained by fitting the oscillatory response with the sum
of four damped oscillators: \textbf{b} central frequencies, \textbf{c}
linewidths, and \textbf{d} spectral weights. $T_{\mathrm{S}}$ is denoted by vertical
dashed lines. The dashed red line in \textbf{c} presents the expected
$T$-dependence of the linewidth of 1.45 THz mode for the case, when damping is
governed by the anharmonic phonon decay.\cite{Cardona} Panels \textbf{e}-\textbf{g} present the zoom-in of the panels \textbf{b}-\textbf{d}, emphasizing the evolution of the parameters across the triclinic transition at $T_{\mathrm{S}}=138$ K. The error bars are obtained from the standard deviation of the least-squared fit. }%
\label{Fig2}%
\end{figure}

Figure \ref{Fig2} presents the results of the analysis of the oscillatory
response. Figure \ref{Fig2}\textbf{a} shows the $T$\textit{-}dependence of the
Fast Fourier Transformation (FFT) spectra in the contour plot, where several
modes up to $\approx6$ THz can be resolved, with the low-T mode frequencies
depicted by red arrows. To analyze the temperature dependence
of the modes' parameters we fit the oscillatory response to a sum of damped
oscillators, $
{\textstyle\sum\nolimits_{i}}
S_{i}\cos\left(  2\pi\widetilde{\nu}_{i}t+\phi_{i}\right)  e^{-\Gamma_{i}t}$.

Figure \ref{Fig2}\textbf{b}-\textbf{d} presents T-dependences of the extracted
mode frequencies $\nu_{i}$ (here $\nu_{i}^{2}=\widetilde{\nu}_{i}^{2}%
+(\Gamma_{i}/2\pi)^{2}$ - see \cite{Zeiger}), dampings $\Gamma_{i} $, and
spectral weights ($S_{i}$) of the four dominant modes (see also Supplementary
Information). Noteworthy, all these low frequency modes are observed up to $\approx150$ K, well above $T_{\mathrm{S}} = 138$ K and $T_{\mathrm{S}'} = 142$ K. While their spectral weights are dramatically reduced upon increasing the temperature through $T_{\mathrm{S}}$,
their frequencies and linewidths remain nearly constant through $T_{\mathrm{S}}$ and $T_{\mathrm{S}'}$.

\begin{figure}[pth]
\centerline{\includegraphics[width=80mm]{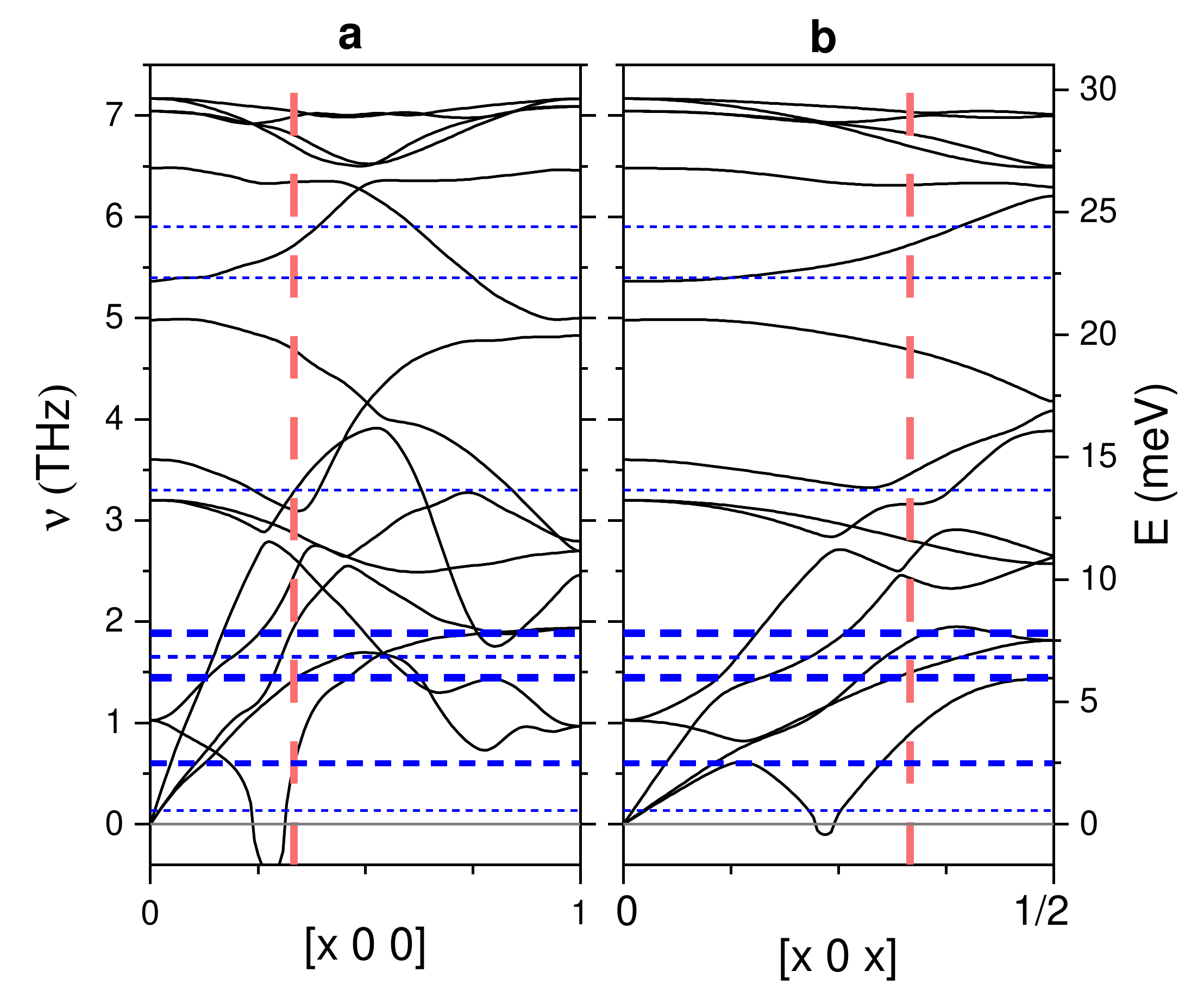}}\caption{\textbf{Phonon dispersion calculation of the high temperature tetragonal structure.} Phonon dispersion along the \textbf{a} [100] and \textbf{b} [101] directions. The dashed red vertical line in
\textbf{a} signifies the CDW wave-vectors of the I-CDW while the line in
\textbf{b} corresponds to the CDW wave-vectors of the C-CDW order. The dashed
horizontal lines indicate the low-temperature frequencies of the observed
modes. Note that calculations show an instability in an optical branch quite
close to the critical wavevector of the I-CDW (see also Methods).}%
\label{Fig3}
\end{figure}

In Figure \ref{Fig3} we present the result of the phonon dispersion
calculations for the high temperature tetragonal structure. None of the
experimentally observed low frequency modes matches the calculated \textbf{q}$=0$  mode
frequencies. Therefore, and based on their $T$- and $F$-dependence, discussed
below, we attribute these modes to collective amplitude modes of the CDW
order.\cite{Scha10,Scha14,rice,Levanyuk,Thomson} In particular, we argue that
these low-temperature \textbf{q}$=0$  amplitude modes are a result of linear (or higher
order\cite{Scha10,Scha14}) coupling of the underlying electronic modulation
with phonons at the wavevector \textbf{q}$_{\mathrm{CDW}}$ (or $n\cdot$ \textbf{q}$_{\mathrm{CDW}}$ for the $n$-th
order coupling\cite{Scha10,Scha14}) of the high-$T$ phase. Within this
scenario,\cite{Scha10,Scha14,rice,Levanyuk,Thomson,Schubert} the low-$T$ frequencies of
amplitude modes should be comparable to frequencies of normal state phonons at
\textbf{q}$_{\mathrm{CDW}}$ (or $n\cdot$ \textbf{q}$_{\mathrm{CDW}}$ for the higher order coupling), with
renormalizations depending on the coupling strengths. Moreover, $T$%
-dependences of modes' parameters $\nu_{i}$, $\Gamma_{i}$ and $S_{i}$ should
reflect the temperature variation of the underlying electronic order
parameter.\cite{Scha10,Scha14,rice,Levanyuk,Thomson,Schubert} 

The first support for the assignment of these modes to amplitude modes follows
from calculations of the phonon dispersion, presented in Fig. \ref{Fig3}. Note
that, since these modes appear already above $T_{S}$, their frequencies must
be compared to phonon dispersion calculations in the high temperature
tetragonal phase. Figure \ref{Fig3} presents the calculated phonon dispersion
in the [100] and [101] directions, along which the modulation of the I-CDW and
C-CDW, respectively, is observed. In Figure \ref{Fig3}, frequencies of the
experimentally observed modes are denoted by the dashed horizontal lines (the
line thicknesses reflect the modes' strengths).

Indeed, the frequencies of strong 1.45 THz and 1.9 THz modes match
surprisingly well with the calculated phonon frequencies at the I-CDW modulation
wavevector (given by the vertical dashed line in Figure \ref{Fig3}\textbf{a}),
supporting the linear-coupling scenario. The corresponding (calculated)
frequencies of phonons at the C-CDW wavevector, shown in Figure \ref{Fig3}%
\textbf{b}, are quite similar. As shown in Fig. \ref{Fig2}, both modes display
a pronounced softening upon increasing temperature, much as the dominant
amplitude modes in the prototype quasi-1D CDW system K$_{0.3}$MoO$_{3}%
$,\cite{Scha10,Scha14} as well as dramatic drop in their spectral weights at
high temperatures.\cite{Scha10} Finally, the particular $T$-dependence of
$\Gamma\ $for the 1.45 THz mode clearly cannot be described by an anharmonic
phonon decay model, given by $\Gamma(\omega,T)=\Gamma_{0}+\Gamma
_{1}(1+2/{e^{{h\nu}/2k_{B}T}-1}).$\cite{Cardona} 
Instead, the behavior is similar to prototype CDW systems, where damping is roughly inversely proportional to the order parameter.\cite{Scha10,Scha14} 

Given the fact that the structural transition at $T_{\mathrm{S}}$ is of the first order, such a strong T-dependence of frequencies and dampings at $T<T_{\mathrm{S}}$ may sound surprising. However, as amplitude modes are a result of coupling between the electronic order and phonons at the CDW wavevector,\cite{Scha10,Scha14,Levanyuk} the $T$-dependence of the mode frequencies and dampings reflect the $T$-dependence of the
electronic order parameter.\cite{Scha10,Scha14} Indeed, the $T$-dependence of
PLD\cite{XRD} as well as of the charge/orbital order\cite{Merz} do display a pronounced $T$-dependence within the C-CDW phase.

A strongly damped mode at 0.6 THz also matches the frequency of the calculated
high-temperature optical phonon at \textbf{q}$_{\mathrm{I-CDW}}$. We note, however, that the
calculations imply this phonon to have an instability near \textbf{q}$_{\mathrm{I-CDW}}$, thus
the matching frequencies should be taken with a grain of salt. The extracted
mode frequency does show a pronounced softening (Fig. \ref{Fig2}\textbf{b}),
though large damping and rapidly decreasing spectral weight result in a large
scatter of the extracted parameters at high temperatures. We further note the
anomalous reduction in damping of the 0.6 THz mode upon increasing the
temperature (Fig. \ref{Fig2}\textbf{c}). Such a behavior has not been observed
in conventional Peierls CDW systems,\cite{Scha10,Scha14} and may reflect the
unconventional nature of the CDW order in this system. We note, that phonon
broadening upon cooling was observed for selected modes in Fe$_{1+y}$%
Te$_{1-x}$Se$_{x}$ \cite{Loidl,LeTacon} and NaFe$_{1-x}$Co$_{x}$As \cite{Um}
above and/or below the respective structural phase transitions. Several interpretation have been put forward for these anomalous anharmonic behaviors, that can have distinct origins.\cite{Loidl,LeTacon,Um}

A weak narrow mode at 1.65 THz is also observed, which does not seem to have a
high temperature phonon counterpart at the \textbf{q}$_{\mathrm{I-CDW}}$. Its low spectral weight
may reflect the higher-order coupling nature of this mode.

Finally, several much weaker modes are also observed (see Figure
\ref{Fig2}\textbf{a}). Comparison with phonon calculations suggest 3.3 THz and 5.4 THz modes are likely regular \textbf{q}$=0$ phonons, the 5.9 THz mode could also be the amplitude collective mode, while the nature of 0.17 THz mode is unclear (see Supplementary Information for further discussion). We note that, as the pump-probe technique is mostly sensitive to A$_{g}$
symmetry modes that couple directly to carrier density \cite{Stevens,Zeiger},
the stronger the coupling to the electronic system, the larger the spectral
weight of the mode. Correspondingly, in time-resolved experiments the spectral
weights of amplitude modes are much higher than regular \textbf{q}$=0$ phonons.

\subsection*{Overdamped modes in BaNi$_{2}$As$_{2}$}

\begin{figure}[pth]
\centerline{\includegraphics[width=90mm]{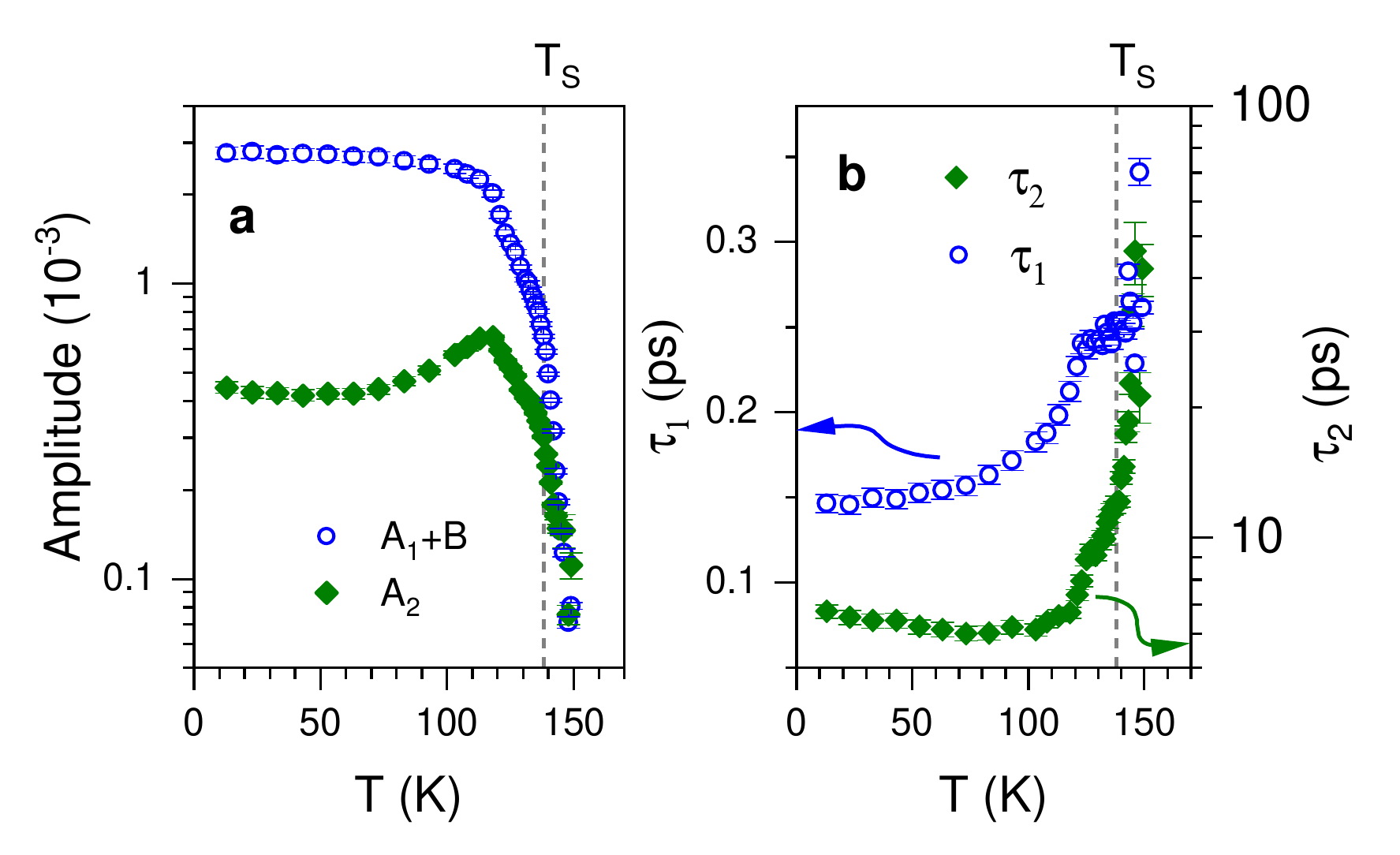}}\caption{\textbf{Extracted fit parameters of the overdamped components.} Temperature
dependence of \textbf{a} amplitudes and \textbf{b} relaxation times, obtained by fitting reflectivity transients using
Eq. (\ref{Eq1}). The error bars are the standard deviation of the least-squared fit.}%
\label{Fig4}%
\end{figure}

Further support for the CDW order in BaNi$_{2}$As$_{2}$\cite{XRD,Merz} is
provided by the $T$-dependence of overdamped components. Figure \ref{Fig4}%
\textbf{a} presents the $T$-dependence of signal amplitudes $A_{1}+B$, which corresponds to the peak value, and $A_{2}$
extracted by fitting the transient reflectivity data using Eq.(\ref{Eq1}). In
CDW systems the fast decay process with $\tau_{1}$ has been attributed to
an overdamped (collective) response of the CDW condensate,\cite{Scha10,Scha14}
while the slower process ($A_{2}$,$\tau_{2}$) has been associated to
incoherently excited collective modes.\cite{Scha14} As both are related to the
CDW order, their amplitudes should reflect this. Indeed, both components are
strongly reduced at high temperatures, with a pronounced change in slope in the
vicinity of $T_{\mathrm{S}}$ - see Figure \ref{Fig4}\textbf{a}. Component $A_{2}$ displays a maximum well below $T_{\mathrm{S}}$, similar to the observation in K$_{0.3}$MoO$_3$.\cite{BBFirst} Above $\approx150$ K the reflectivity transient shows a characteristic metallic response, with fast
decay on the 100 fs timescale.

The evolution of timescales $\tau_{1}$ and $\tau_{2}$ is shown in Figure
\ref{Fig4}\textbf{b}. In the C-CDW phase, up to $\approx110-120$ K, the two
timescales show qualitatively similar dependence as in prototype 1D
CDWs:\cite{Scha10,Scha14,TSI} $\tau_{1}$ increases with increasing temperature
while $\tau_{2}$ decreases.\cite{Scha10,Scha14,TSI} As $\tau_{1}$ is inversely
proportional to the CDW strength,\cite{Scha10,Scha14} its $T$-dependence is
consistent with the observed softening of the amplitude modes. Its increase with increasing temperature is, however, not as pronounced as in CDW systems with continuous phase transitions, where timescales can change by an order of
magnitude when gap is closing in a mean-field
fashion.\cite{TaSTaSe,Scha10,TSI,Scha14} From about 130 K $\tau_{1}$ remains
nearly constant up to $\approx150$ K. On the other hand, for $T\gtrsim120$ K
$\tau_{2}$ displays a pronounced increase, though the uncertainties of the
extracted parameters start to diverge as signals start to faint. Importantly,
all of the observables seem to evolve continuously through $T_{\mathrm{S}}$, despite
the pronounced changes in the electronic and structural properties that are
observed, \textit{e.g.,} in the c-axis transport\cite{Sefat} or the optical
conductivity.\cite{WangPRB,Wang}

\subsection*{Excitation density dependence}

Valuable information about the nature of CDW order can be obtained from
studies of dynamics as a function of excitation fluence, $F$. In conventional
Peierls CDW systems a saturation of the amplitude of the overdamped response
is commonly observed at excitation fluences of the order of 0.1-1 mJ cm$^{-2}%
$.\cite{Tome09,TSI,Yusupov,Stojchevska} The corresponding absorbed energy
density, at which saturation is reached, is comparable to the electronic part of the CDW condensation
energy.\cite{Tome09,Stojchevska} Similarly, the spectral weights of amplitude
modes saturate at this saturation fluence. The modes are still observed up
to excitation densities at which the absorbed energy density reaches the
energy density required to heat up the excited volume up to the CDW transition
temperature.\cite{Tome09} The reason for this is an ultrafast recovery of the
electronic order on a timescale $\tau_{1}$, which is faster than the
collective modes' periods.\cite{Tome09}

\begin{figure}[pth]
\centerline{\includegraphics[width=120mm]{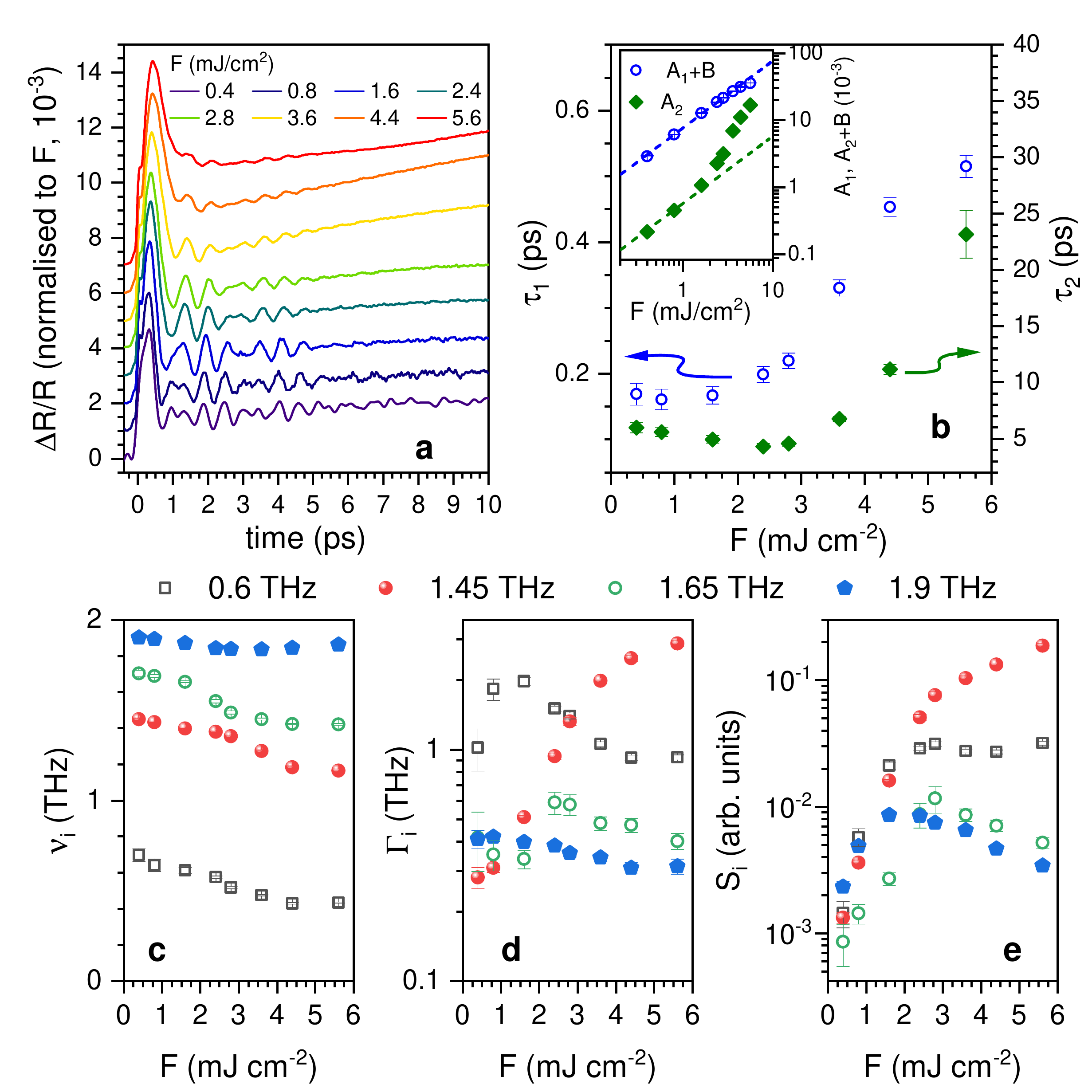}}\caption{ \textbf{Excitation
density dependence of collective dynamics recorded at 10 K.} \textbf{a}
Reflectivity transients normalized to excitation fluence. \textbf{b} The extracted relaxation timescales $\tau_{1}$ and $\tau_{2}$ as a function of excitation fluence, $F$. Inset presents the F-dependence of amplitudes, with dashed lines presenting linear fits. \textbf{c-e} $F$-dependence of the collective mode parameters
$\nu_{i}$, $\Gamma_{i}$, $S_{i}$. The error bars are obtained from the standard deviation of the least-squared fit. }%
\label{Fig5}%
\end{figure}

We performed $F-$dependence study at 10 K base temperature, with $F$ varied
between $0.4$ and $5.6$ mJ cm$^{-2}$. The reflectivity transients are presented in Fig.
\ref{Fig5}\textbf{a}. Unlike in prototype CDWs, no
saturation of the fast overdamped response is observed up to the highest $F$ (inset to Fig. \ref{Fig5}\textbf{b}). The absence of spectroscopic signature of the CDW induced gap in BaNi$_{2}$As$_{2}$\cite{Wang} suggest that most of
the Fermi surface remains unaffected by the CDW order. Thus, the photoexcited
carriers can effectively transfer their energy to the lattice,\cite{Obergfell}
just as in the high-$T$ metallic phase. Nevertheless, the fact that the
excitation densities used here do exceed saturation densities in conventional CDW
systems by over an order of magnitude suggests an unconventional mechanism
driving the CDW in BaNi$_{2}$As$_{2}$. We note that signal $A_2$ displays a super-linear dependence for $F > 2$ mJ cm$^{-2}$. 

Figure \ref{Fig5}\textbf{b} presents $\tau_{1}(F)$ and $\tau_{2}(F)$ for the
data recorded at 10 K. Qualitatively, the $F$-dependence of the two timescales
resembles their temperature dependence, similar to observations in Peierls
CDW systems.\cite{Tome09} Since $\tau_{1}$ reflects the recovery of the
electronic part of the order parameter, $\Delta$, and follows $\tau_{1}%
\propto1/\Delta$,\cite{Scha10,Scha14} this observation supports a continuous
suppression of the electronic order with increasing $F$. However, in Ni-122 no discontinuous drop in $\tau_{1}(F)$ is observed up to the highest fluences. In K$_{0.3}$MoO$_{3}$\cite{Tome09} such a drop in $\tau_{1}(F)$ is observed at the fluence corresponding to the full suppression of the electronic order.

Figure \ref{Fig5}\textbf{c}-\textbf{e} presents the $F$-dependence of
the extracted amplitude mode parameters. A softening upon increasing the fluence is
observed for all four modes  (Fig. \ref{Fig5}\textbf{c}). However, above $\approx3$ mJ cm$^{-2}$ the values reach a plateau. Such an unusual behavior is not observed in Peierls CDWs\cite{Scha10,Scha14} and may hold clues to the
interplay between the periodic lattice distortion and the underlying 
electronic instability. An indication of suppression of the underlying
electronic order is observed also as saturation of spectral weights of some of
the amplitude modes near $F\approx3$ mJ cm$^{-2}$, see Fig. \ref{Fig5}%
\textbf{e.} On the other hand, the mode at 1.45 THz, which is the most similar to main modes in K$_{0.3}$MoO$_{3}$, shows no such saturation up to the highest fluences. While the observed anomalies seen near $F\approx3$ mJ cm$^{-2}$ may be linked to the underlying microscopic mechanism of CDW order in Ni-122, one could also speculate the anomalies  may be related to the photoinduced suppression of commensurability.

To put the observed robustness of the CDW against optical excitation into
perspective, we note that $F=1$ mJ cm$^{-2}$ corresponds to the absorbed energy
density of about 180 J cm$^{-3}$ (110 meV per formula unit). Assuming rapid
thermalization between electrons and the lattice, and no other energy decay
channels, the resulting temperature of the excited sample volume would reach
$\approx160$ K (see also Supplementary Information). However, with high conductivity
also along the c-axis \cite{Sefat} and the estimated electronic mean free path
on 7 nm \cite{Kurita}, transport of hot carriers into the bulk on the
(sub)picosecond timescale cannot be excluded. Nevertheless, the fact that even
at 5.6 mJ cm$^{-2}$ (0.6 eV per formula unit) the CDW order has not collapsed,
underscores an unconventional CDW order in in BaNi$_{2}$As$_{2}$%
.\cite{XRD,Merz}

\section*{Conclusion}

Our results clearly demonstrate the existence of CDW collective modes in BaNi$_{2}$As$_{2}$, which appear well above the triclinic transition. At temperatures well below the triclinic transition, the modes show qualitatively similar temperature dependence as in extensively studied prototype 1D CDW K$_{0.3}$MoO$_{3}$ \cite{Scha10,Scha14,Dominko1}. For temperatures above $\approx 130$ K, however, only spectral weights of the modes get suppressed, while their central frequencies and dampings remain largely constant up to $\approx 150$ K. This provides an important insight into the relation between the CDW order and the structural phase transitions. While the XRD data\cite{XRD,Lee21,Merz} clearly show two distinct modulations above and below T$_{\mathrm{S}}$, the collective modes show no detectable discontinuity of their frequencies and dampings at $T_{\mathrm{S}}$ (nor at $T_{\mathrm{S}'}$). This suggests that the C-CDW evolves from the I-CDW$_{1}$ by gaining additional periodicity along the c-axis. The sequence of phase transitions,
with orthorhombicity accompanying the appearance of unidirectional I-CDW$_{1}$,\cite{Merz}
may suggest charge-order driven nematicity in BaNi$_{2}$As$_{2}$. Moreover,
the fact that $T_{\mathrm{S}}$ coincides with I-CDW$_{1}-$C-CDW transition may in fact
support the idea of structural phase transition being mediated by the
stabilization of the CDW order. In an alternative scenario, the lock-in CDW
supports the triclinic phase, which otherwise competes with the tetragonal/orthorhombic
one, as suggested by the strong reduction of the c/a ratio when entering the
triclinic phase, and the first order nature of the transition. The change of
the CDW modulation vector is then triggered by the underlying structural change.

While the $T$-dependence of collective mode dynamics roughly follows the behavior seen
in conventional Peierls CDWs, implying the existence of an underlying
electronic instability, the resilience of the electronic CDW order against perturbations
suggests an unconventional mechanism. Recent photoemission
data\cite{Noda} suggest the band reconstruction to be consistent with the
proposed orbitally driven Peierls instability.\cite{Khomskii1,Khomskii} Such a
scenario is further supported by the finding of Ni-Ni dimers.\cite{Merz}
Moreover, also a third type of commensurate CDW order, with \textbf{q}$_{\mathrm{CDW}}%
=(1/2,0,1/2)$ was observed in Ba$_{1−x}$Sr$_{x}$Ni$_{2}$As$_{2}$ for $x>0.4$.\cite{Lee21} Thus, systematic doping and pressure dependent studies of collective modes may provide valuable additional clues to the underlying microscopic interactions.

Our findings suggest an intimate relation between charge-order and
structural instabilities in a Ni-122 system and imply an unconventional origin
of the electronic instability, likely associated to orbital
ordering.\cite{Khomskii1,Khomskii} Together with the observed doping
dependence of superconducting critical temperature,\cite{Kudo} the results
provide important input for theoretical models addressing the interplay
between high-temperature superconductivity to a close proximity of a competing
electronic instability.

\section*{Methods}

\subsection*{Single crystals of BaNi$_{2}$As$_{2}$}

Single crystals of BaNi$_{2}$As$_{2}$ with typical dimensions 2$\times $2$%
\times$0.5 mm$^{3}$ were grown by self-flux method similar to reported
literature \cite{Kudo,Sefat}. Crystals were mechanically freed from the flux
and characterized using X-ray diffraction and energy-dispersive x-ray
spectroscopy (EDX). The samples were cleaved along the $a-b$ plane before mounting
into an optical cryostat and were kept in vacuum during the measurements.

\subsection*{Experimental set-up}

A commercial 300 kHz Ti:Sapphire amplifier producing 50 fs laser pulses at $%
\lambda$ = 800 nm (photon energy of 1.55 eV) was used as a source of both
pump and probe pulse trains. The beams were at near normal incidence, with
polarizations at 90 degrees with respect to each other to reduce noise. The
fluence was varied between 0.1 - 5 mJ cm$^{-2}$ while the probe fluence was
about 30 $\mu$J cm$^{-2}$. The induced changes of an in-plane reflectivity
($R$) were recorded utilizing a fast-scan technique, enabling high
signal-to-noise level \cite{Scha10}. Continuous sample heating was
experimentally estimated by performing temperature scans at different
excitation densities. For the excitation fluence of 0.4 mJ cm$^{-2}$ the
continuous laser heating near $T_{\mathrm{S}}$ results in the probed sample volume
about 3 K higher than the base temperature; the continuous laser heating has
been taken into account when plotting the temperature dependent data.

\subsection*{Phonon dispersion calculation}

Lattice dynamics properties for the high-temperature tetragonal structure
were calculated using the linear response or density-functional perturbation
theory (DFPT) implemented in the mixed-basis pseudo-potential method \cite%
{Heid,Louie,Meyer}. The electron-ion interaction is described by
norm-conserving pseudo-potentials, which were constructed following the
descriptions of Hamann, Schl\"{u}ter, Chiang \cite{HSC} for Ba and
Vanderbilt \cite{Vanderbilt} for Ni and As, respectively. Semi-core states
Ba-5$p$, Ni-3$s$, Ni-3$p$ were included in the valence space.

In the mixed-basis approach, valence states are expanded in a combination of
plane waves and local functions at atomic sites, which allows an efficient
description of more localized components of the valence states. Here, plane
waves with a cut-off for the kinetic energy of 22 Ry and local functions of $%
p$,$d$ type for Ba and $s$, $p$, $d$ type for Ni, respectively, were
employed. Brillouin-zone integration was performed by sampling a tetragonal $%
16\times16\times8$ k-point mesh in conjunction with a Gaussian broadening of
50 meV. The exchange-correlation functional was represented by the
general-gradient approximation in the Perdew-Burke-Ernzerhof form \cite{PBE}.

Phonon dispersions along the [100] and [101] directions of the tetragonal
structure, shown in Figure \ref{Fig3}, were obtained by Fourier-interpolation of
dynamical matrices calculated by DFPT on a tetragonal $8\times8\times2$
mesh. A denser 16 x 1 x 1 mesh was used to better resolve the position of the instability in the [100] direction. Structural parameters were taken from room-temperature measurements
after \cite{Merz}.

We find similar phonon branches as in a previous work \cite{Subedi}, with
one notable exception: there occurs an instability in an optical branch near
the critical wavevector of the I-CDW of (0.28,0,0), which was missed in the
previous study. This instability only shows up, when a sufficiently dense
mesh for the calculation of the dynamical matrices is used, and is easily
overlooked otherwise.

\section*{Data availability}

All relevant data are available from the authors. 

\bibliography{BNA_publist}

\section*{Acknowledgments}

This work was funded by the Deutsche Forschungsgemeinschaft (DFG, German
Research Foundation) - TRR 288 - 422213477 (projects B03 and B08). The
contribution from M.M. was supported by the Karlsruhe Nano Micro Facility
(KNMF). R.H. acknowledges support by the state of Baden-W\"{u}rttemberg
through bwHPC.

\section*{Author contributions}

A.R.P, V.G., A.M., T.D. performed ultrafast optical spectroscopy measurements. A.A.H. grew the single crystal samples and performed EDX characterization. M.M. performed XRD measurements. Y.L. performed Raman studies. R.H. performed phonon dispersion calculations. V.G., A.R.P., and JD analyzed the data. V.G., A.R.P. and J.D. wrote the manuscript with contributions from all coauthors. J.D. coordinated the project.

\section*{Competing interests}

The authors declare no competing interests.

\end{document}